\documentclass[sigconf,nonacm,balance=false]{acmart}

\usepackage{enumitem}

\setlist[itemize,enumerate]{
  topsep=3pt,       
  partopsep=0pt,    
  itemsep=1pt,      
  parsep=0pt       
}

\setlist[itemize,1]{leftmargin=1.3em,labelsep=0.4em}
\setlist[enumerate,1]{leftmargin=1.6em,labelsep=0.4em}

\usepackage{multirow}
\usepackage{graphicx}
\usepackage{subcaption}
\usepackage{array}
\usepackage{cuted}

\usepackage{placeins}
\setcopyright{none}
\begin{document}

\title[PFA: Attention Can Be Progressively Focused in Long-Sequence Generative Recommendation]{PFA: Attention Can Be Progressively Focused in Long-Sequence Generative Recommendation}
\author{Pengjie Zhou}
\orcid{0009-0000-9691-8243}
\authornote{These authors contributed equally to this work.}
\affiliation{%
  \institution{Alibaba International Digital Commerce Group}
  \city{Beijing}
  \country{China}
}
\email{sy2442115@buaa.edu.cn}

\author{Weiqin Yang}
\orcid{0000-0002-5750-5515}
\authornotemark[1]
\affiliation{%
  \institution{Zhejiang University}
  \city{Hangzhou}
  \country{China}
}
\email{tinysnow@zju.edu.cn}

\author{Haibo Xing}
\orcid{0009-0006-5786-7627}
\authornotemark[1]
\affiliation{%
  \institution{Alibaba International Digital Commerce Group}
  \city{Hangzhou}
  \country{China}
}
\email{xinghaibo.xhb@alibaba-inc.com}

\author{Hao Deng}
\orcid{0009-0002-6335-7405}
\affiliation{%
  \institution{Alibaba International Digital Commerce Group}
  \city{Beijing}
  \country{China}
}
\email{denghao.deng@alibaba-inc.com}

\author{Jinxin Hu}
\orcid{0000-0002-7252-5207}
\authornote{Corresponding author.}
\affiliation{%
  \institution{Alibaba International Digital Commerce Group}
  \city{Beijing}
  \country{China}
}
\email{jinxin.hjx@alibaba-inc.com}

\renewcommand{\shortauthors}{Zhou et al.}

\begin{abstract}
Long user behavioral sequences improve the performance of generative recommendation (GR), but computing full attention over long histories during autoregressive beam search decoding incurs substantial online inference latency. 
To accelerate long-sequence decoding, existing approaches typically rely on static history compression or dynamic top-$k$ sparse attention; however, the former risks losing critical semantics, whereas the latter incurs significant overhead from repeated token selection and non-contiguous memory access.
In this work, we thoroughly analyze attention patterns during GR decoding and reveal two key properties: (i) \textbf{progressive attention focusing}, where attention progressively focuses on a small subset of critical historical tokens; and (ii) \textbf{cross-step ranking consistency}, where their importance rankings remain relatively stable across subsequent decoding steps.

Motivated by these observations, we propose \textbf{P}rogressively \textbf{F}ocused \textbf{A}ttention (\textbf{PFA}), an efficient sparse decoding framework for long-sequence GR. PFA retains full attention at the initial decoding step to explore global historical preferences. It then establishes a shared importance ranking through cross-beam weighted aggregation and reorders the key-value cache of the selected tokens into a contiguous memory layout. Subsequent decoding steps reuse this ranking and geometrically reduce the token budget, preserving critical semantics while lowering computational and memory access overhead. Extensive experiments on two public benchmarks and a large-scale industrial dataset demonstrate that PFA outperforms state-of-the-art baselines. A 7-day online A/B test on a large-scale e-commerce advertising platform further confirms that PFA delivers substantial business gains while maintaining low inference latency. Our code is publicly available at \url{https://anonymous.4open.science/r/PFA-review-088D}.
\end{abstract}

\ccsdesc[500]{Information systems~Recommender systems}
\keywords{Generative Recommendation, Long-Sequence Modeling, Sparse Attention}

\maketitle

\FloatBarrier
\section{Introduction}
\label{sec:introduction}

\begin{figure*}[!t]
  \centering
  \begin{subfigure}[t]{0.485\textwidth}
    \centering
    \includegraphics[height=0.62\linewidth]{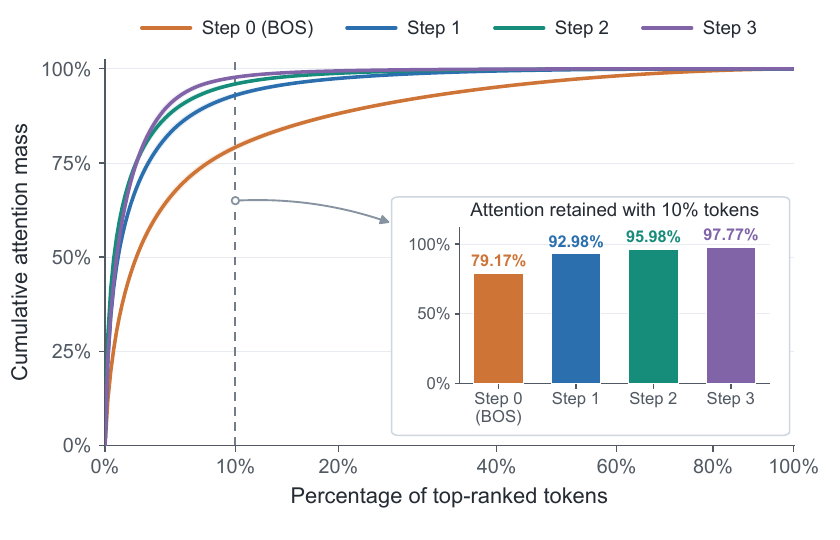}
    \caption{Progressive attention focusing}
    \label{fig:attention-concentration}
  \end{subfigure}
  \hfill
  \begin{subfigure}[t]{0.485\textwidth}
    \centering
    \includegraphics[height=0.62\linewidth]{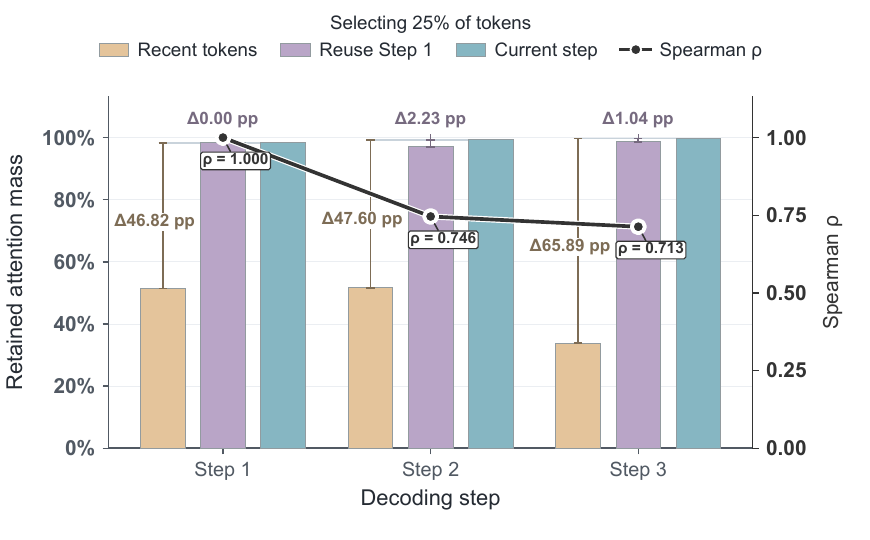}
    \caption{Cross-step ranking consistency}
    \label{fig:cross-step-reuse}
  \end{subfigure}
  \caption{\textbf{Progressive Attention Focusing and Cross-Step Ranking Consistency.}
\textbf{(a)} Cumulative attention curves over historical tokens sorted in descending order of attention weight at each decoding step. The top 10\% of tokens capture 79.17\% of the attention mass at Step~0 and 97.77\% at Step~3, illustrating progressive attention focusing.
\textbf{(b)} Retained attention mass (bars) and cross-step Spearman rank correlations with Step~1 (line). Reusing the top-$k$ historical tokens selected at Step~1 ($k/N=25\%$) retains 97.03\% and 98.62\% of the attention mass at Steps~2 and~3, respectively—only 2.23 and 1.04 percentage points (pp) below current-step top-$k$ selection. By contrast, selecting the most recent $k$ tokens yields substantially larger gaps of 47.60 and 65.89~pp relative to the same reference. For the fixed token set selected at Step~1, the Spearman rank correlations between its Step~1 importance ranking and its rankings at Steps~2 and~3 are 0.746 and 0.713, respectively, supporting cross-step token reuse.}

 \Description{Pilot experiments supporting the motivation of our approach.}
  \label{fig:pilot-industrial}
\end{figure*}

Generative recommendation (GR) has recently emerged as a promising paradigm for recommender systems~\cite{li2025survey,rajput2023recommender}. It represents user behavioral histories as sequences of discrete semantic tokens and reformulates candidate item retrieval as autoregressive generation via beam search over a codebook space. Recent studies demonstrate that GR substantially enhances the modeling of complex behavioral semantics and user preferences, driving its widespread adoption in industrial recommender systems~\cite{zheng2025ega,chen2025onesearch,xing2026reg4rec,fu2026forge}.

However, in industrial recommender systems, GR often faces a severe online inference efficiency bottleneck caused by extremely long user behavioral sequences~\cite{yang2025earn}. To retrieve a list of candidate items, a generative recommender decodes multiple independent beams in parallel. At each decoding step, every beam computes full attention between its current query and the entire behavioral history~\cite{sun2025xgr}, causing the decoding cost to scale linearly with history length. In practice, scaling the history length by $4\times$ leads to a $2.46\times$ increase in inference latency. This prohibitive overhead makes it challenging for GR to meet the stringent latency constraints of large-scale recommender systems with long user interaction histories.

Existing approaches to reducing the decoding cost of full attention over long behavioral histories fall into two main categories:
\begin{itemize}
  \item \textbf{Static history compression.} This approach aggregates historical interactions via predefined rules or lightweight models prior to decoding, thereby shortening the history accessed by the decoder~\cite{chai2025longer,li2023blip,zhou2025onerec,liu2026accelerating,zhan2026rastp}. However, because compression is decoupled from decoding, it struggles to capture attention patterns that vary across decoding steps (see Figure~\ref{fig:pilot-industrial}(\subref{fig:attention-concentration})). Consequently, critical semantic information required at specific steps may be irreversibly lost.
  \item \textbf{Dynamic top-$k$ sparse attention.} This method addresses the above limitation by dynamically retrieving the top-$k$ most relevant historical tokens based on the current query at each decoding step~\cite{cao2026coarse,zhou2026gems,yi2026dualgr}. While this approach adapts to step-wise attention patterns and preserves step-specific critical semantics, its efficiency gains remain limited under beam search. Specifically, because each beam independently selects top-$k$ tokens at every step, it induces frequent non-contiguous key-value (KV) cache accesses~\cite{yuan2025native}, severely degrading inference efficiency (e.g., a $2.36\times$ slowdown on the Industry dataset; see Table~\ref{tab:ablation}).
\end{itemize}

These limitations necessitate a decoding mechanism that preserves critical semantics while improving inference efficiency in long-sequence GR. To this end, we systematically analyze how attention over the historical sequence changes across decoding steps (see Figure~\ref{fig:pilot-industrial}). Our analysis reveals that \textbf{attention naturally becomes progressively focused as decoding proceeds}. As shown in Figure~\ref{fig:pilot-industrial}(\subref{fig:attention-concentration}), the model initially attends to a broad range of historical tokens to capture diverse user preferences; in subsequent steps, the majority of the attention mass is focused on progressively fewer tokens (e.g., starting from decoding Step~1, merely 10\% of the tokens account for 92.98\% of the attention mass).  
More importantly, \textbf{these critical tokens exhibit strong ranking consistency across decoding steps}. Specifically, the top-$k$ token rankings between subsequent steps and Step~1 achieve an average Spearman rank correlation of 0.729 (Figure~\ref{fig:pilot-industrial}(\subref{fig:cross-step-reuse})).
Consequently, the model can preserve critical semantics without repeatedly searching the entire history for relevant tokens; instead, it can establish a shared importance ranking early in decoding and progressively narrow the attention scope.

Motivated by these observed attention patterns, we propose \textbf{Progressively Focused Attention (PFA)}, an efficient sparse decoding framework for long-sequence GR. Specifically, PFA retains full attention at the initial decoding step (Step~0) to thoroughly explore global historical preferences. Subsequently, at Step~1, PFA performs cross-beam importance-based token selection. It uses cross-beam weighted aggregation of the Step~1 attention distributions to compute an importance score for each historical token. From the resulting shared importance ranking, PFA selects the top-$k$ tokens to form an initial critical historical token set $\mathcal I_1$ shared across beams. This set preserves essential semantics: just 25\% of the historical tokens cover an average of \textbf{97.83\% of the attention mass} in subsequent decoding steps (see Figure~\ref{fig:pilot-industrial}(\subref{fig:cross-step-reuse})). For subsequent decoding steps ($t \ge 2$), PFA directly reuses this critical token set while geometrically reducing the number of attended historical tokens, further lowering decoding overhead without redundant importance retrieval. Furthermore, PFA supports sharing KV cache subsets across beams and decoding steps, mitigating non-contiguous memory access overhead and significantly accelerating decoding.

Across two public benchmarks and a large-scale industrial dataset, PFA demonstrates superior recommendation quality and inference efficiency. Offline evaluations show that PFA achieves average relative improvements of \textbf{5.75\%} in Recall@10 and \textbf{7.18\%} in NDCG@10 over the strongest baseline for each dataset and metric (Table~\ref{tab:overall-performance}). On the Industry dataset, it achieves a \textbf{1.45$\times$} inference speedup over the Full Attention baseline, while improving Recall@10 by \textbf{3.16\%} (Table~\ref{tab:ablation}). Furthermore, in a 7-day online A/B test in a production e-commerce advertising system, PFA increases advertising revenue (REV) by \textbf{3.76\%} over the production baseline, with an online response time (RT) of \textbf{11\,ms} (Table~\ref{tab:online_ab}).

Our main contributions are summarized as follows:

\begin{itemize}
  \item We systematically characterize how attention over historical sequences changes during autoregressive decoding in long-sequence GR, revealing both progressive attention focusing and cross-step ranking consistency.
  \item We propose PFA, which establishes a shared importance ranking of critical historical tokens early in decoding via cross-beam weighted aggregation. By reusing this ranking and geometrically reducing the token budget in subsequent decoding steps, PFA preserves essential semantics while significantly accelerating inference.
  \item We conduct extensive evaluations on multiple public benchmarks, a large-scale industrial dataset, and a live production recommender system, demonstrating the superior efficiency and effectiveness of PFA.
\end{itemize}

\FloatBarrier
\section{Preliminaries}
\label{sec:preliminaries}

\begin{figure*}[!t]
  \centering
    \includegraphics[width=0.99\textwidth]{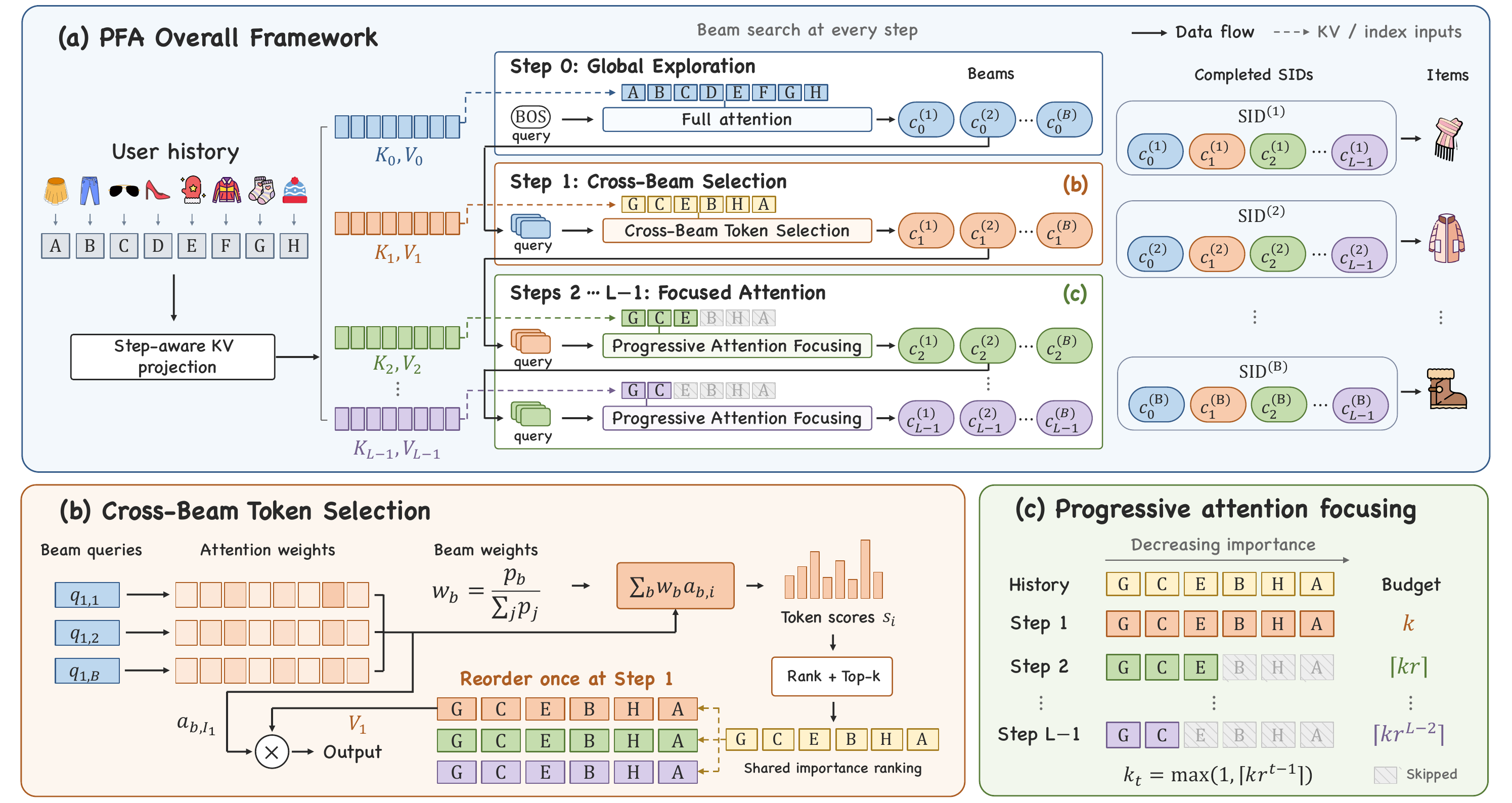}%
  \caption{Overview of the PFA framework.
\textbf{(a) Overall framework.} PFA precomputes step-specific historical KV representations via step-aware KV projection. Step~0 attends to the complete user history to capture global preferences. Step~1 constructs a cross-beam shared token importance ranking; subsequent steps ($t \geq 2$) reuse this ranking and progressively shrink the token budget as beam search generates candidate items.
\textbf{(b) Cross-beam token selection.} At Step~1, PFA aggregates attention distributions across beams, weighted by their normalized generation probabilities, to select the top-$k$ historical tokens. Corresponding step-specific KV representations are reordered by descending importance (e.g., G, C, E, B, H, A) into compact, contiguous caches shared across beams.
\textbf{(c) Geometric budget reduction.} For $t \geq 2$, PFA geometrically decays the token budget to $k_t$ and attends only to the first $k_t$ tokens in the reordered KV cache at step $t$. Reusing this ranking enables direct prefix truncation, eliminating redundant full-history token selection and reducing both attention computation and KV memory traffic.}

  \Description{Framework diagram of PFA.}
  \label{fig:method-overview}
\end{figure*}

\subsection{Generative Recommendation}
\label{sec:gr}

GR reformulates conventional item retrieval as the autoregressive generation of discrete semantic IDs (SIDs)~\cite{liu2025generative}. Specifically, each item $v$ is represented as a sequence of $L$ discrete semantic tokens, denoted as $\mathrm{SID}_v=(c_0,\ldots,c_{L-1})$. Given a user $u$ with an interaction history $\mathcal{H}_u=(v_1,\ldots,v_N)$ of length $N$, the model encodes $\mathcal{H}_u$ into a hidden-state matrix $\mathbf{H}\in\mathbb{R}^{N\times d}$, where $d$ is the hidden dimension. At decoding step $t$, the model predicts $c_t$ conditioned on the encoded history $\mathbf{H}$ and the generated prefix $c_{<t}=(c_0,\ldots,c_{t-1})$. The target SID probability factorizes as follows:

\begin{equation}
\label{eq:gr}
P(\mathrm{SID}_v\mid\mathbf{H})
=\prod_{t=0}^{L-1}P(c_t\mid c_{<t},\mathbf{H}).
\end{equation}

\subsection{Beam Search}
\label{sec:beam-search}
Because greedy decoding commits to a single token at each sequential step, it is prone to suboptimal local decisions and can easily miss globally high-probability items. Moreover, recommendation tasks typically require multiple candidate items, motivating the use of beam search to maintain several high-probability SID prefixes~\cite{lin2025efficient}.

Given a beam width $B$, beam search expands the active candidate prefixes at each decoding step and retains the top-$B$ prefixes with the highest cumulative log-probabilities. Formally, the score of a candidate prefix $c_{<t}$ of length $t$ is defined as:

\begin{equation}
\label{eq:beam-score}
S(c_{<t})
=\sum_{j=0}^{t-1}\log P(c_j\mid c_{<j},\mathbf{H}).
\end{equation}

After $L$ decoding steps, the generated valid SIDs are mapped back to candidate items. However, because each beam independently predicts the next token conditioned on its own decoding state, beam search substantially amplifies both the computational overhead and memory access costs of autoregressive decoding.

\subsection{Efficiency Bottlenecks in Long-Sequence Generative Recommendation}
\label{sec:bottlenecks}

\paragraph{Inference cost for long sequences} At each autoregressive decoding step, every beam attends to the user's encoded interaction history $\mathbf{H}$, resulting in an overall attention complexity of $\mathcal{O}(BLNd)$. Although different beams can share the  KV cache of the historical sequence, this cache must still be repeatedly read from memory at every decoding step. Thus, both the attention computation and KV memory access overhead scale linearly with the history length $N$. Coupled with multi-step decoding and beam search, these factors establish attention over long histories as a primary inference bottleneck in GR.

\paragraph{Limitations of static history compression} Static history compression mitigates decoding costs by merging tokens or aggregating features prior to decoding, condensing the full history representations into a shorter sequence. Representative methods such as LONGER~\cite{chai2025longer} and Q-Former~\cite{li2023blip,zhou2025onerec} follow this paradigm, which can be formulated as:
\begin{equation}
\label{eq:compression}
\widetilde{\mathbf{H}}=f_{\mathrm{comp}}(\mathbf{H})\in\mathbb{R}^{M\times d},\quad M<N,
\end{equation}
where $f_{\mathrm{comp}}(\cdot)$ denotes the compression function and $M$ is the compressed sequence length. During decoding, $\widetilde{\mathbf{H}}$ replaces the full history $\mathbf{H}$ as the context for attention at every step, reducing both the attention computation and KV memory access overhead. However, because compression is performed once before decoding and remains static across decoding states, it cannot adapt to evolving information needs across steps and risks discarding fine-grained semantics required by later tokens. Once lost, this information cannot be recovered during subsequent decoding, ultimately degrading recommendation quality.

\paragraph{Efficiency bottlenecks of dynamic top-$k$ sparse attention} To prevent premature information loss, dynamic top-$k$ sparse attention leverages the current decoding query at each step to retrieve the top-$k$ most relevant historical tokens for attention. While this strategy adapts to step-specific information needs, it requires each beam to independently perform token selection at every decoding step, introducing substantial selection overhead. Moreover, it necessitates retaining the full-history KV cache in memory. Because the selected tokens are typically scattered across non-contiguous memory locations, gathering their KV states incurs irregular memory access and high data movement overhead. Consequently, the theoretical FLOPs reduction from attending to fewer tokens rarely translates into actual wall-clock inference speedup.

\section{Methodology}
\label{sec:method}

\subsection{Design Motivation}
\label{sec:motivation}

Unlike general natural language generation, where attention patterns often shift dynamically across tokens, GR exhibits two distinctive attention properties during autoregressive decoding: \emph{progressive attention focusing} and \emph{cross-step ranking consistency}. First, attention becomes progressively focused on a small subset of critical historical tokens, motivating a progressively shrinking token budget as decoding proceeds. Second, critical tokens maintain consistent importance rankings across steps, and critical tokens identified early continue to capture most of the attention mass in subsequent steps, justifying the reuse of early rankings. However, conventional dynamic top-$k$ sparse attention methods overlook these GR-specific properties; by repeatedly retrieving over the full history and maintaining a static token budget at every step, they incur substantial redundant computation. In contrast, our key insight is that the model can establish a shared importance ranking of historical tokens early in decoding and progressively prune the tokens participating in attention. Leveraging these properties, we propose \textbf{PFA}, which reuses the shared importance ranking and geometrically reduces the token budget to preserve critical semantics while accelerating decoding.

\subsection{Progressively Focused Attention}
\label{sec:pfa}

As illustrated in Figure~\ref{fig:method-overview}, PFA organizes autoregressive decoding into three stages: global preference exploration, cross-beam importance-based token selection, and geometric token budget reduction.

\paragraph{Stage 1: Global Preference Exploration.}
At the initial decoding step $t=0$, PFA performs full attention to allow the decoder to access the complete user history:
\begin{equation}
\label{eq:global-attention}
\mathbf o_0=
\operatorname{Attention}(\mathbf q_0,\mathbf K_0,\mathbf V_0),
\end{equation}
where the initial query $\mathbf q_0$ is derived from the beginning-of-sequence token $\langle\mathrm{BOS}\rangle$, and $\mathbf K_0$ and $\mathbf V_0$ denote the full historical KV representations at step $t=0$. Based on the attention output $\mathbf o_0$, the model predicts the first semantic token and retains the top-$B$ candidate paths via beam search. These beams yield distinct semantic prefixes, which subsequently guide the identification of critical historical tokens across diverse user interests.

\paragraph{Stage 2: Cross-Beam Importance-Based Token Selection.}
At decoding step $t=1$, the distinct semantic prefixes across different beams capture diverse facets of the user's historical interests. PFA assigns each historical token an aggregated importance score through cross-beam weighted aggregation. It then constructs a shared importance ranking and selects an initial critical token set shared across beams.

Specifically, for beam $b$, the attention weight assigned to historical token $i$ is given by:
\begin{equation}
\label{eq:token-attention}
a_{b,i}=
\operatorname{Softmax}
\left(
\frac{\mathbf q_{1,b}\mathbf K_{1}^{\top}}
{\sqrt{d_k}}
\right)_i,
    \end{equation}
where $\mathbf q_{1,b}$ is the query of beam $b$ at step $t=1$, $\mathbf K_{1}$ is the full historical key matrix for this step, and $d_k$ is the key dimension. In multi-head attention, $a_{b,i}$ can be averaged across all heads.

Next, PFA performs cross-beam weighted aggregation of these attention distributions using each beam's normalized generation probability:
\begin{equation}
\label{eq:beam-aggregation}
w_b=\frac{p_b}{\sum_{j=1}^{B}p_j},
\qquad
s_i=\sum_{b=1}^{B}w_ba_{b,i},
\end{equation}
where $p_b$ is the generation probability of beam $b$, and $s_i$ denotes the aggregated importance score of historical token $i$. Sorting the historical tokens by $s_i$ in descending order yields a shared importance ranking $\boldsymbol{\pi}$. PFA selects the top-$k$ tokens to form the initial critical token set $\mathcal I_1$, which is shared across beams:
\begin{equation}
\label{eq:ranking}
\boldsymbol{\pi}=\operatorname{argsort}_{i}(s_i),
\qquad
\mathcal I_1=\{\pi_1,\ldots,\pi_k\}.
\end{equation}

According to this ranking, PFA reorders the precomputed historical KV representations for subsequent decoding steps, retaining only the top-$k$ tokens to form a compact, contiguous KV cache. At step $t=1$, PFA reuses the previously computed attention weights,
renormalizes them over the critical token set $\mathcal I_1$,
and computes the attention output using the corresponding value vectors. All beams share this reordered KV cache, whose contiguous memory layout enables zero-overhead prefix truncation in later decoding steps.

\paragraph{Stage 3: Geometric Token Budget Reduction.}
Leveraging the shared importance ranking from Stage~2, PFA starts from the initial critical token set $\mathcal I_1$ and geometrically shrinks the historical token budget as decoding proceeds:
\begin{equation}
\label{eq:token-budget}
k_t=\max\left(1,\left\lceil kr^{t-1}\right\rceil\right),
\quad t\geq1,\; 0<r\leq1,
\end{equation}
where $k$ is the initial token budget at $t=1$, $r$ is the geometric retention ratio, and $\lceil\cdot\rceil$ denotes the ceiling operator.
Accordingly, the critical token set at decoding step $t$ is defined as:
\begin{equation}
\label{eq:critical-set}
\mathcal I_t=\{\pi_1,\ldots,\pi_{k_t}\}.
\end{equation}

Reusing the shared importance ranking $\boldsymbol{\pi}$ across steps naturally produces nested critical token sets satisfying $\mathcal I_{t+1}\subseteq\mathcal I_t$. Consequently, for $t\geq2$, each beam $b$ performs attention exclusively over the truncated critical token set $\mathcal I_t$:
\begin{equation}
\label{eq:sparse-attention}
\mathbf o_{t,b}=
\operatorname{Attention}
\left(
\mathbf q_{t,b},
\mathbf K_{t,\mathcal I_t},
\mathbf V_{t,\mathcal I_t}
\right),
\end{equation}
where $\mathbf K_{t,\mathcal I_t}$ and $\mathbf V_{t,\mathcal I_t}$ denote the key and value representations sliced at the shared indices $\mathcal I_t$ for step $t$. By coupling reuse of the shared importance ranking with geometric token budget reduction, PFA eliminates repeated full-history retrieval and progressively reduces attention complexity across decoding steps.

\paragraph{Step-Aware KV Projection.}
Although different decoding steps attend to nested critical token sets, predicting hierarchical semantic tokens at different steps requires distinct semantic information from the user history. To enhance step-specific expressiveness, PFA assigns independent KV projection parameters to each decoding step and precomputes the corresponding historical KV caches:
\begin{equation}
\label{eq:step-aware-kv}
\mathbf K_t = \mathbf H\mathbf W_K^{(t)}, \quad \mathbf V_t = \mathbf H\mathbf W_V^{(t)}, \quad t=0,\ldots,L-1,
\end{equation}
where $\mathbf W_K^{(t)}$ and $\mathbf W_V^{(t)}$ are the step-specific key and value projection matrices for step $t$. Projecting the reused historical tokens into distinct semantic subspaces across steps effectively prevents representational homogenization and mitigates potential representation collapse~\cite{yang2026pitfall}.
\subsection{Efficiency and Effectiveness Analysis}
\label{sec:efficiency}

\paragraph{Decoding Efficiency.}
PFA substantially reduces attention computation during decoding via cross-beam importance-based token selection and geometric token budget reduction. Table~\ref{tab:attention_complexity} compares the theoretical computational costs and speedups of Full Attention,
Dynamic Top-$k$ Sparse Attention, and PFA. Under a typical industrial configuration, PFA achieves theoretical attention speedups of $2.65\times$ over full attention and $1.99\times$ over dynamic top-$k$ sparse attention. These gains stem from two factors: reusing a shared importance ranking across beams and decoding steps eliminates redundant token selection, and geometrically decaying the token budget progressively narrows the attention scope. This advantage becomes more pronounced as the initial selection ratio $k/N$ decreases or the SID length $L$ increases. Moreover, dynamic top-$k$ sparse attention requires per-beam top-$k$ selection and non-contiguous KV gather operations at each decoding step, incurring substantial indexing and memory access overhead. By sharing token indices across beams and maintaining a contiguous KV cache, PFA avoids redundant selection and irregular memory access, further improving practical decoding efficiency.

\begin{table}[!htbp]
  \centering
  \caption{Theoretical attention computation and speedup comparison.
  We set $N=512$, $B=100$, $L=4$, $d=256$,
  $k=N/2=256$, and $r=0.5$.
  MACs (M) denotes millions of multiply-accumulate operations.
  Speedup is defined as the ratio of each baseline's
  attention MACs to those of PFA.}
  \label{tab:attention_complexity}

  \footnotesize
  \setlength{\tabcolsep}{3pt}
  \renewcommand{\arraystretch}{1.2}

  \resizebox{\columnwidth}{!}{%
    \begin{tabular}{lccc}
      \toprule
      Method & Attention MACs & MACs (M) & Speedup \\
      \midrule
      Full Attention
      & $2[1+B(L-1)]Nd$
      & 78.91
      & $\mathbf{2.65\times}$ \\

      Dynamic Top-$k$ Sparse Attention
      & $[1+B(L-1)](N+k)d$
      & 59.18
      & $\mathbf{1.99\times}$ \\
      \midrule

      \textbf{PFA (Ours)}
      & $\left[(B+2)N+Bk+2B\sum_{t=2}^{L-1}k_t\right]d$
      & \textbf{29.75}
      & -- \\
      \bottomrule
    \end{tabular}%
  }
\end{table}

\paragraph{Semantic Preservation.}
Beyond acceleration, PFA preserves recommendation quality through three key designs. First, retaining all historical tokens at the initial decoding step preserves full context for global preference exploration, avoiding the premature information loss caused by static compression. Second, our cross-beam token selection and geometric budget reduction align with the progressive attention focusing inherent to GR decoding, retaining salient semantics while filtering out irrelevant noise. Finally, the step-aware KV projection maps reused tokens into step-specific semantic subspaces, enhancing stage-wise representational capacity for hierarchical SID prediction.

Overall, PFA strikes an effective balance between critical semantic preservation and decoding efficiency in long-sequence GR.

\FloatBarrier
\section{Experiments}
\label{sec:experiments}

We comprehensively evaluate the recommendation quality and inference efficiency of PFA on two public benchmarks and one large-scale industrial dataset, and further validate its practical effectiveness through online A/B testing. Our experiments are designed to answer the following five research questions (RQs):

\begin{itemize}
  \item \textbf{RQ1:} How does PFA perform compared with representative baseline methods in terms of recommendation quality?
  \item \textbf{RQ2:} How does each core component of PFA contribute to overall recommendation quality and inference efficiency?
  \item \textbf{RQ3:} How does the timing of shared index construction affect recommendation quality and inference latency?
  \item \textbf{RQ4:} How sensitive is PFA to the initial token budget $k$ and its geometric retention ratio $r$?
  \item \textbf{RQ5:} Does PFA achieve significant online improvements in a live production recommender system?
\end{itemize}

\subsection{Experimental Setup}
\label{sec:setup}

\paragraph{Datasets} We evaluate PFA on three datasets: Kindle Store, Tmall, and Industry, covering e-book recommendation, click-based e-commerce recommendation, and e-commerce advertising recommendation, respectively. 
Kindle Store is derived from Amazon Reviews 2023~\cite{hou2026bridging}.
Tmall is constructed from click interactions in the IJCAI-15 Repeat Buyers Prediction Dataset~\cite{tianchi2018tmall}.
Industry is collected from a large-scale e-commerce advertising platform.
Table~\ref{tab:dataset_statistics} summarizes the statistics of the datasets.

\begin{table}[!htbp]
    \centering
    \caption{Statistics of the datasets.}
    \label{tab:dataset_statistics}
    \renewcommand{\arraystretch}{1.1}
    \begin{tabular}{cccc}
        \toprule
        Statistic & Kindle Store & Tmall & Industry \\
        \midrule
        \#Users & 5,801 & 19,093 & 35,154,135 \\
        \#Items & 98,013 & 204,470 & 48,106,880 \\
        \#Interactions & 1,590,326 & 5,100,848 & 626,020,176 \\
        Seq. Len. ($N$) & 128 & 128 & 512 \\
        \bottomrule
    \end{tabular}
\end{table}

\paragraph{Data Preprocessing and Splitting} For Kindle Store, we treat interactions with ratings of at least 4 as positive feedback and retain active users with at least 128 positive records. For Tmall and Industry, we construct interaction sequences from user clicks, retaining only the earliest interaction for repeated user-item pairs. Across all datasets, each user's interactions are sorted chronologically and partitioned into training, validation, and test phases at an approximate $90\%:5\%:5\%$ ratio. We construct prediction instances using a sliding window under a rolling-history evaluation protocol, where each input sequence comprises up to $N$ observed behaviors strictly preceding the target interaction. For the public datasets, we iteratively filter out items with fewer than 5 occurrences in the training instances, remove users without valid sequences, and apply deterministic sampling to form the final splits.

\paragraph{Baselines} We compare PFA against eight representative baselines across four categories: (1) conventional sequential recommenders, \textbf{SASRec}~\cite{kang2018self} and \textbf{HSTU}~\cite{zhai2024actions}; (2) the standard generative recommender, \textbf{TIGER}~\cite{rajput2023recommender}; (3) general-purpose static history compression methods for long sequences, \textbf{LONGER}~\cite{chai2025longer}  and \textbf{Q-Former}~\cite{li2023blip,zhou2025onerec}; and (4) long-sequence generative recommenders, \textbf{RASTP}~\cite{zhan2026rastp}, \textbf{DualGR}~\cite{yi2026dualgr}, and \textbf{GEMs}~\cite{zhou2026gems}. For a fair comparison, we enforce an identical attention token budget across all efficient long-sequence methods, whereas the remaining baselines attend to the full user history.

\paragraph{Evaluation Metrics} We evaluate offline recommendation quality using $\text{Recall}@K$ and $\text{NDCG}@K$~\cite{jarvelin2002cumulated} with $K\in\{5, 10\}$, which measure target item retrieval and ranking performance, respectively. For online A/B testing, we assess practical business impact using REV and click-through rate (CTR).

\paragraph{Implementation Details} All models are implemented in PyTorch with an embedding dimension of 256. The generative recommenders adopt a Transformer-based encoder-decoder architecture comprising 2 encoder layers, 2 decoder layers, and 8 attention heads. 
The prediction head is a two-layer MLP with input/output dimensions of $256$, a hidden dimension of $640$, and a SwiGLU activation function.
The SID codebook configuration is set to $[128, 128, 128, 128]$ for Kindle Store and Tmall, and $[300, 300, 300, 300]$ for Industry. We optimize all models using AdamW with a learning rate of $0.001$ and a weight decay of $0.01$. Models are trained for up to 50 epochs with early stopping on the public datasets, and for a single epoch on the Industry dataset.

\paragraph{Efficiency Measurement} We benchmark inference latency across 5 NVIDIA L20 GPUs using BF16 mixed precision, with a per-GPU batch size of 512 and a beam width of 100. Latency measurements span the encoder forward pass and the full autoregressive beam search decoding process, excluding data loading and external embedding lookups. We perform 3 warmup batches and apply CUDA synchronization before and after timing, reporting the average inference latency per batch per GPU (ms/batch).

\subsection{Overall Performance (RQ1)}
\label{sec:overall}

Table~\ref{tab:overall-performance} summarizes the overall recommendation performance of all compared models across the three datasets.

\begin{table*}[!t]
  \centering
  \caption{Overall performance comparison across the three datasets
  in terms of Recall@5, NDCG@5, Recall@10, and NDCG@10.
  The best and second-best results for each metric on each dataset
  are highlighted in \textbf{bold} and \underline{underlined}, respectively.
  \textit{Improv.} denotes the relative improvement of PFA over the strongest baseline.}
  \label{tab:overall-performance}

  \footnotesize
  \setlength{\tabcolsep}{2pt}
  \renewcommand{\arraystretch}{1.15}

  \begin{tabular*}{\textwidth}{
    @{\extracolsep{\fill}}l*{12}{c}@{}
  }
    \toprule
    \multirow{2}{*}{Method}
    & \multicolumn{4}{c}{Kindle Store}
    & \multicolumn{4}{c}{Tmall}
    & \multicolumn{4}{c}{Industry} \\
    \cmidrule(lr){2-5}
    \cmidrule(lr){6-9}
    \cmidrule(lr){10-13}

    & Recall@5 & NDCG@5 & Recall@10 & NDCG@10
    & Recall@5 & NDCG@5 & Recall@10 & NDCG@10
    & Recall@5 & NDCG@5 & Recall@10 & NDCG@10 \\
    \midrule

    SASRec
    & 0.0294 & 0.0214 & 0.0442 & 0.0259
    & 0.0511 & 0.0317 & 0.0978 & 0.0462
    & 0.0724 & 0.0468 & 0.1223 & 0.0623 \\

    HSTU
    & 0.0302 & 0.0228 & 0.0487 & 0.0289
    & 0.0543 & 0.0339 & 0.0994 & 0.0487
    & 0.0792 & 0.0504 & 0.1256 & 0.0655 \\

    TIGER
    & 0.0327 & 0.0235 & 0.0532 & 0.0299
    & 0.0596 & 0.0356 & 0.1043 & 0.0506
    & 0.0845 & 0.0562 & 0.1332 & 0.0722 \\

    LONGER
    & \underline{0.0393}
    & 0.0262
    & \underline{0.0611}
    & \underline{0.0332}
    & 0.0661
    & \underline{0.0414}
    & 0.1132
    & 0.0563
    & \underline{0.1086}
    & 0.0768
    & 0.1565
    & 0.0928 \\

    Q-Former
    & 0.0368 & 0.0246 & 0.0579 & 0.0312
    & 0.0632 & 0.0392 & 0.1103 & 0.0547
    & 0.1072 & 0.0761 & 0.1552 & 0.0913 \\

    RASTP
    & 0.0379 & 0.0257 & 0.0575 & 0.0322
    & 0.0628 & 0.0389 & 0.1096 & 0.0536
    & 0.1056 & 0.0745 & 0.1541 & 0.0904 \\

    DualGR
    & 0.0375 & 0.0252 & 0.0571 & 0.0315
    & 0.0617 & 0.0379 & 0.1075 & 0.0528
    & 0.1048 & 0.0736 & 0.1528 & 0.0886 \\

    GEMs
    & 0.0391
    & \underline{0.0264}
    & 0.0603
    & 0.0331
    & \underline{0.0667}
    & 0.0412
    & \underline{0.1148}
    & \underline{0.0570}
    & 0.1082
    & \underline{0.0782}
    & \underline{0.1581}
    & \underline{0.0947} \\

    \midrule

    \textbf{PFA}
    & \textbf{0.0415}
    & \textbf{0.0278}
    & \textbf{0.0635}
    & \textbf{0.0349}
    & \textbf{0.0710}
    & \textbf{0.0452}
    & \textbf{0.1218}
    & \textbf{0.0619}
    & \textbf{0.1178}
    & \textbf{0.0848}
    & \textbf{0.1695}
    & \textbf{0.1021} \\

    \textit{Improv.}
    & 5.60\% & 5.30\% & 3.93\% & 5.12\%
    & 6.45\% & 9.18\% & 6.10\% & 8.60\%
    & 8.47\% & 8.44\% & 7.21\% & 7.81\% \\

    \bottomrule
  \end{tabular*}
\end{table*}

\paragraph{Recommendation Quality} PFA consistently achieves the best performance across all four metrics on all three datasets. Compared with the strongest baseline for each metric, PFA yields average relative improvements of \textbf{5.75\%} in $\text{Recall}@10$ and \textbf{7.18\%} in $\text{NDCG}@10$ across the three datasets. Under an identical token budget, PFA also substantially outperforms existing static history compression methods, demonstrating its superior ability to preserve historical information relevant to target prediction. This advantage stems from two key designs: first, rather than prematurely truncating or compressing token information before decoding, PFA retains the full contextual semantics throughout encoding and performs cross-beam importance-based token selection during decoding for effective semantic preservation; second, the step-aware KV projection further enriches the representational capacity at each decoding step.

\subsection{Ablation Study (RQ2)}
\label{sec:ablation}

To evaluate the contribution of each core component, we conduct ablation studies on Kindle Store and Industry, with the results summarized in Table~\ref{tab:ablation}. We compare the full PFA model against five variants:

\begin{enumerate}
  \item \textbf{Full Attention:} Applies full-sequence attention at every decoding step without token pruning.
  \item \textbf{Dynamic Top-$k$ Sparse Attention:} Performs top-$k$ token selection independently for each decoding step and each beam.
  \item \textbf{w/o cross-beam sharing:} Constructs and reuses sparse indices independently within each beam rather than sharing them across beams.
  \item \textbf{w/o geometric reduction:} Fixes the token budget to its initial value across all subsequent decoding steps.
  \item \textbf{w/o step-aware KV projection:} Shares a single set of KV projection parameters across all decoding steps.
\end{enumerate}

\begin{table}[!htbp]
  \centering
  \caption{Ablation study of PFA on Kindle Store and Industry.}
  \label{tab:ablation}
  \footnotesize
  \setlength{\tabcolsep}{2pt}
  \renewcommand{\arraystretch}{1.15}
  \begin{tabular*}{\columnwidth}{
    @{\extracolsep{\fill}}
    >{\raggedright\arraybackslash}m{0.42\columnwidth}rrrr@{}
  }
    \toprule
    \multirow{2}{*}{Method}
    & \multicolumn{2}{c}{Kindle Store}
    & \multicolumn{2}{c}{Industry} \\
    \cmidrule(lr){2-3}
    \cmidrule(l){4-5}
    & Recall@10 & Latency & Recall@10 & Latency \\
    \midrule
    Full Attention & 0.0624 & 24 & 0.1643 & 90 \\
    Dynamic Top-$k$ Sparse Attention & 0.0623 & 53 & 0.1638 & 212 \\
    \midrule
    w/o cross-beam sharing
      & 0.0648 & 29 & 0.1706 & 118 \\
    w/o geometric reduction
      & 0.0637 & 20 & 0.1699 & 65 \\
    w/o step-aware KV projection
      & 0.0618 & 16 & 0.1602 & 59 \\
    \midrule
    \textbf{PFA} & 0.0635 & 18 & 0.1695 & 62 \\
    \bottomrule
  \end{tabular*}
\end{table}

\paragraph{Effectiveness of the Overall Architecture}
While Dynamic Top-$k$ Sparse Attention matches the accuracy of Full Attention, per-step and per-beam top-$k$ selection triggers non-contiguous GPU memory accesses, more than doubling inference latency.
By integrating cross-beam sharing, geometric reduction, and step-aware KV projection, PFA simultaneously improves recommendation accuracy and substantially reduces inference latency compared with Full Attention.

\paragraph{Impact of Cross-Beam Sharing and Geometric Reduction} Disabling cross-beam sharing sharply increases inference latency with only marginal accuracy gains, confirming that using shared indices across beams effectively eliminates redundant top-$k$ sorting and scattered KV reads. Similarly, removing geometric reduction increases latency while yielding negligible differences in accuracy. Together, these two mechanisms prune redundant computation along the beam and decoding-step dimensions with minimal impact on recommendation quality.

\paragraph{Effect of Step-Aware KV Projection} Removing step-aware KV projection significantly degrades recommendation accuracy while yielding only minimal latency savings. This confirms that assigning dedicated KV projection parameters to each decoding step effectively captures stage-specific historical semantics with negligible computational overhead.

\subsection{Timing of Shared Index Construction (RQ3)}
\label{sec:timing}

To examine how the timing of shared index construction affects model performance, we compare four strategies on the Industry dataset, triggering construction from Step~0 to Step~3. All strategies share the same initial token budget $k$: they apply full-sequence attention prior to shared index construction, and subsequently reuse the shared indices under the same geometric token budget reduction schedule. We evaluate both Recall@10 and end-to-end inference latency.

\begin{table}[!htbp]
    \centering
    \caption{Impact of shared index construction timing.
$\dagger$ denotes the default construction step used in the main experiments.}
    \label{tab:selection_timing}
    \small
    \setlength{\tabcolsep}{8pt} 
    \renewcommand{\arraystretch}{1.15}
    \begin{tabular}{lcccc}
        \toprule
        \multirow{2}{*}{Metric} & \multicolumn{4}{c}{Construction Step} \\
        \cmidrule(lr){2-5}
        & Step~0 & Step~1$^\dagger$ & Step~2 & Step~3 \\
        \midrule
        Recall@10 & 0.1632 & 0.1695 & 0.1701 & 0.1704 \\
        Latency   & 55     & 62     & 94     & 103 \\
        \bottomrule
    \end{tabular}
\end{table}

\paragraph{Results} As shown in Table~\ref{tab:selection_timing}, constructing the shared indices at Step~1 strikes a favorable balance between recommendation accuracy and inference efficiency. Compared to Step~0, Step~1 improves Recall@10 by 0.0063 (3.86\% relatively) while increasing end-to-end inference latency by only 7 ms/batch. This aligns with our preliminary observations: attention distributions are relatively diffuse at Step~0, but rapidly become focused on a small subset of relevant historical behaviors once the first semantic token is generated, enabling more accurate selection of the  critical token set at Step~1. Moreover, compared to deferring construction to Step~2 or Step~3, Step~1 requires fewer full-sequence attention operations, avoiding substantial latency overhead for marginal accuracy gains. Consequently, PFA constructs the shared indices at Step~1 by default.

\subsection{Hyperparameter Study (RQ4)}
\label{sec:hyperparameters}

On the Industry dataset, we investigate the sensitivity of PFA to the initial token budget $k$ and the geometric retention ratio $r$, varying one hyperparameter while keeping all other settings fixed. Table~\ref{tab:hyperparameter} summarizes the results.

\begin{table}[!htbp]
    \centering
    \caption{Hyperparameter sensitivity analysis.
We fix $r=0.50$ when varying $k$, and $k=256$ when varying $r$.
$\dagger$ denotes the default setting used in the main experiments.}

    \label{tab:hyperparameter}
    \small
    \setlength{\tabcolsep}{5pt} 
    \renewcommand{\arraystretch}{1.15} 
    
    \begin{tabular}[t]{ccc} 
        \toprule
        \multicolumn{3}{c}{Parameter $k$} \\
        \cmidrule(lr){1-3}
        Value & Recall@10 & Latency \\
        \midrule
        384 & 0.1704 & 78 \\
        320 & 0.1702 & 69 \\
        256\rlap{$^\dagger$} & 0.1695 & 62 \\
        192 & 0.1692 & 57 \\
        128 & 0.1648 & 52 \\
        \bottomrule
    \end{tabular}%
    \hspace{12pt}
    \begin{tabular}[t]{ccc} 
        \toprule
        \multicolumn{3}{c}{Parameter $r$} \\
        \cmidrule(lr){1-3}
        Value & Recall@10 & Latency \\
        \midrule
        1.00 & 0.1699 & 65 \\[0.85ex] 
        0.75 & 0.1697 & 63 \\[0.85ex]
        0.50\rlap{$^\dagger$} & 0.1695 & 62 \\[0.85ex]
        0.25 & 0.1664 & 60 \\[0.85ex]
        \bottomrule
    \end{tabular}
\end{table}

\paragraph{Initial token budget $k$} We vary $k$ across $\{384, 320, 256, 192, 128\}$. Reducing $k$ from 384 to 256 substantially lowers inference latency while maintaining comparable recommendation accuracy. However, further decreasing $k$ ( especially down to 128) leads to a noticeable drop in Recall@10 with diminishing latency reductions. These results indicate that moderately pruning historical tokens effectively eliminates redundant computation, whereas overly aggressive compression discards critical user preferences. Consequently, we set $k=256$ by default to balance recommendation accuracy and inference efficiency.

\paragraph{Geometric retention ratio $r$} We vary $r$ across $\{1.00, 0.75, 0.50, 0.25\}$, where $r=1.00$ disables geometric token budget reduction. As $r$ decreases from 1.00 to 0.50, inference latency steadily declines while Recall@10 remains robust. Reducing $r$ further to 0.25 markedly degrades recommendation accuracy while yielding minimal latency savings. This confirms that moderate geometric reduction effectively curtails redundant memory access in later decoding steps without losing essential historical context. We therefore adopt $r=0.50$ by default.

\subsection{Online A/B Testing (RQ5)}
\label{sec:online}
To validate the effectiveness of PFA in a real-world production environment, we deploy it on a large-scale e-commerce advertising platform and conduct a 7-day online A/B test. The control group serves the baseline production recommender with user behavior sequences truncated to 128, whereas the treatment group integrates PFA to extend the sequence length to 512. Each group is randomly allocated 10\% of live production traffic.

As shown in Table~\ref{tab:online_ab}, the treatment group achieves relative improvements of 3.76\% in REV, 0.34\% in CTR, and 2.56\% in gross merchandise value (GMV) over the control group. Regarding system efficiency, PFA maintains an online RT of \textbf{11~ms}, which is notably lower than the \textbf{15~ms} incurred by directly feeding the full 512-item sequence into the computation graph. These results demonstrate that PFA enables generative recommenders to effectively leverage long user behavior sequences for substantial business gains while simultaneously accelerating online inference.

\begin{table}[!htbp]
\centering
\caption{Results of the online A/B test.}
\label{tab:online_ab}
\small
\setlength{\tabcolsep}{6pt}
\renewcommand{\arraystretch}{1.15}
\begin{tabular*}{\columnwidth}{
    @{\extracolsep{\fill}}lcccc@{}
}
\toprule
Method & REV & CTR & GMV & RT (ms) \\
\midrule
PFA & +3.76\% & +0.34\% & +2.56\% & 11 \\
\bottomrule
\end{tabular*}
\end{table}

\FloatBarrier
\section{Related Work}
\label{sec:related-work}
\paragraph{Generative Recommendation.}
Generative recommendation reformulates recommendation as conditional sequence generation~\cite{geng2022recommendation}, demonstrating substantial potential in large-scale industrial systems~\cite{zhou2025onerec,zhai2024actions,he2026plum}.
A typical paradigm maps items to SIDs composed of discrete codewords, and then autoregressively generates target identifiers conditioned on user context~\cite{rajput2023recommender,wang2024eager,liu2026mmgrec,xu2026mmq}.
Subsequent studies align item representations with the generation process by enriching codewords with content semantics and collaborative signals~\cite{wang2024learnable,xiao2025unger,liu2025discrec}, or by jointly optimizing tokenizers and recommenders toward recommendation objectives~\cite{liu2025generative,bai2026bi,wang2026pit,fu2026differentiable,li2026unigrec}.
However, as interaction histories accumulate, accurately capturing long-term preferences while maintaining efficient autoregressive inference remains a pressing bottleneck.

\paragraph{Long-Sequence Modeling in Recommendation.}
Traditional discriminative recommenders leverage target candidates as queries to retrieve or aggregate relevant historical behaviors~\cite{pi2020search,chen2021end,cao2022sampling,chang2023twin}, or reduce computational overhead via feature compression, sparse access, and external memory~\cite{chai2025longer,yuan2026hisac,ma2026blossomrec,tan2021dynamic}.
Lacking explicit target candidates during inference, generative recommenders must preserve diverse user interests while constraining history access throughout autoregressive search.
Consequently, one line of work focuses on history pre-compression and state maintenance.
For instance, CAUSE aggregates long-term behaviors by category alongside recent interactions within HSTU~\cite{liu2026accelerating}.
To achieve internal denoising and multi-granular fusion, RASTP prunes uninformative history tokens at intermediate Transformer layers based on semantic saliency and attention centrality~\cite{zhan2026rastp}; GRACE integrates block compression, query-aware block selection, coarse journey tokens, and recent sliding windows via gating mechanisms~\cite{ma2025grace}.
For streaming updates, Rec2PM jointly encodes existing preference memory and incoming behavior segments, recurrently maintaining memory tokens for subsequent predictions~\cite{chen2026recurrent}.
Another line of research retrieves historical behaviors on demand using intermediate decoding signals, such as partially generated prefixes or current-step queries.
DualGR predicts a coarse SID to retrieve behaviors from the corresponding bucket for fine-grained decoding~\cite{yi2026dualgr}; similarly, GLASS summarizes long-term interests for first-codeword prediction, and then uses this codeword to retrieve relevant history and gate it into subsequent decoding steps~\cite{cao2026coarse}.
GEMs dynamically selects top-$k$ mid-term KV entries via a lightweight query-based indexer, combining them with recent interactions and pre-compressed lifelong representations~\cite{zhou2026gems}.
While these methods mitigate computational costs through compression and single-step retrieval, they largely overlook the multi-step nature of autoregressive decoding when modeling long-sequence user histories in GR---specifically, how to efficiently share and reuse historical context across decoding steps and candidate beams.

\paragraph{Efficient Sequence Modeling for Large Models.}
Research on large language models has established several sequence-modeling paradigms to accelerate long-context inference.
Common strategies include compressing long contexts into shorter text or summary vectors~\cite{pan2024llmlingua,chen2025dast,chu2026kwai}; recurrently updating hidden states via linear attention or state-space models to bypass full-history traversal~\cite{katharopoulos2020transformers,yang2025gated,gu2023mamba,dao2024transformers}; and reducing attention complexity through local windows and content-aware token or block selection~\cite{beltagy2020longformer,jiang2024minference,lai2025flexprefill,fan2026flashprefill,fan2026flashprefill2,tang2024quest,yuan2025native,lu2026moba}.
As a mainstream paradigm, top-$k$ dynamic sparse attention substantially reduces attention FLOPs, yet online history scoring and selection often emerge as new latency bottlenecks~\cite{yang2025lserve,bai2026indexcache,xu2026hisa}.
To eliminate redundant selection overhead, recent studies explore index and state reuse across multiple dimensions.
For example, IndexCache and ChunkKV share retrieval indices across layers~\cite{bai2026indexcache,liu2026chunkkv}---a strategy adopted in frontier models such as GLM-5.2~\cite{glm5team2026glm5vibecodingagentic} and DeepSeek-V4.1-Flash~\cite{xu2026deepseek}---while other works share selected context across decoding steps to amortize selection costs~\cite{yang2025lserve,takbir2026flexicache}.
This reuse philosophy has recently been extended to generative recommendation: OneLA shares historical recurrent states of linear attention across beams to accelerate wide-beam search~\cite{yang2026onela}.
Nevertheless, for standard softmax attention mechanisms that remain widely adopted, designing efficient long-sequence modeling tailored to the specific decoding structure of generative recommendation remains largely underexplored.

\FloatBarrier
\section{Conclusion}
\label{sec:conclusion}

In this paper, we present PFA, an efficient sparse decoding framework for long-sequence generative recommendation. Specifically, we identify two key properties of historical attention during autoregressive decoding: progressive attention focusing and cross-step ranking consistency. Motivated by these observations, PFA retains full attention for initial preference exploration and constructs a shared token importance ranking through cross-beam weighted aggregation. By reusing this ranking and geometrically reducing the token budget in subsequent steps, PFA preserves critical historical semantics while minimizing redundant token selection overhead and KV cache access. Furthermore, we incorporate a step-aware KV projection to capture the distinct semantic requirements across different decoding steps. Extensive offline experiments on real-world datasets and online A/B testing in a commercial advertising system demonstrate that PFA simultaneously improves recommendation quality and accelerates inference, validating its practical value in supporting longer user histories. In future work, we aim to explore adaptive token budget allocation to better accommodate diverse user histories and dynamic decoding states.

\bibliographystyle{ACM-Reference-Format}
\bibliography{references}

\appendix
\section{Additional Attention Analysis}
\label{app:attention-analysis}

To further verify the generalizability of our findings, Figure~\ref{fig:pilot-amazon}
illustrates that the Kindle Store dataset mirrors the progressive attention focusing and cross-step ranking consistency
previously observed on Industry.

\begin{strip}
  \captionsetup{type=figure}
  \centering
  \begin{subfigure}[t]{0.485\textwidth}
    \centering
    \includegraphics[height=0.62\linewidth]{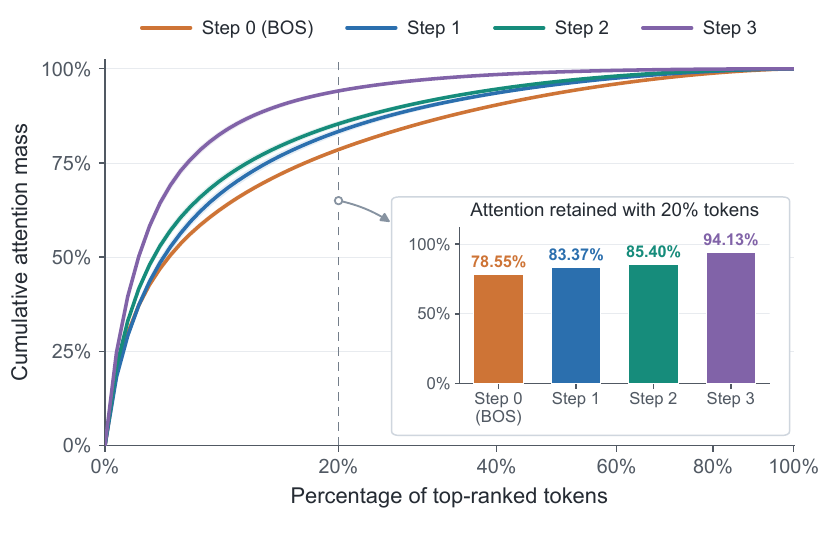}
    \caption{Progressive attention focusing}
    \label{fig:attention-concentration-amazon}
  \end{subfigure}
  \hfill
  \begin{subfigure}[t]{0.485\textwidth}
    \centering
    \includegraphics[height=0.62\linewidth]{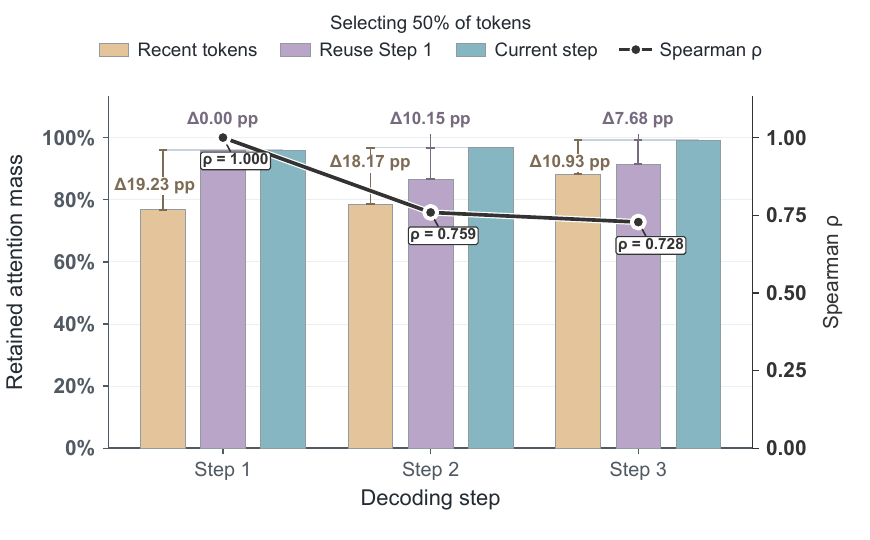}
    \caption{Cross-step ranking consistency}
    \label{fig:cross-step-reuse-amazon}
  \end{subfigure}
  \caption{\textbf{Progressive Attention Focusing and Cross-Step Ranking Consistency on Kindle Store.}
    \textbf{(a)} Cumulative attention curves of sorted historical tokens.
    The attention mass accounted for by the top 20\% tokens surges from 78.55\% at Step~0
    to 94.13\% at Step~3, evidencing a clear trend of progressive attention focusing.
    \textbf{(b)} Retained attention mass (bars) and Spearman rank correlations
    with Step~1 (line) across subsequent steps. Directly reusing the top 50\% tokens from Step~1
    successfully captures 86.57\% and 91.51\% of the total attention mass at Steps~2 and~3,
    respectively, while preserving high rank correlations (0.759 and 0.728).
    This pronounced consistency strongly justifies cross-step token reuse.}
  \label{fig:pilot-amazon}
  \Description{Two panels illustrating progressive attention focusing and cross-step
    ranking consistency on the Kindle Store dataset. Panel (a) plots the cumulative attention
    mass across decoding steps. Panel (b) compares the attention mass retained by
    recent-token selection, Step-1 token reuse, and current-step selection,
    alongside Spearman rank correlations relative to Step~1.}
\end{strip}

\par\null

\end{document}